\documentstyle[pra,epsf,aps]{revtex}
\begin{document}
\title{A Minimal Model for Dilatonic Gravity}
\author{P.~P.~Fiziev \thanks{ E-mail:\,\, fiziev@phys.uni-sofia.bg}}
\address{Department of Theoretical Physics,
 University of Sofia, \\ 5 James Bourchier Boulevard, BG-1164, Sofia, Bulgaria
}
\maketitle
\begin{abstract}
We study a new minimal scalar-tensor model of gravity with
Brans-Dicke factor $\omega(\Phi)\equiv 0$ and cosmological factor
$\Pi(\Phi)$. The constraints on $\Pi(\Phi)$ from known
gravitational experiments are derived. We show that almost any
time evolution of the scale factor in a homogeneous isotropic Universe
can be obtained via a properly chosen  $\Pi(\Phi)$ and discuss the
general properties of models of this type.
\end{abstract}
%%%%%%%%%%%%%%%%%%%%%%%%%%%%%%%%%%%%%%%%%%%%%%%%%%%%%%%%%%%%%%%%%%%
%\draft
\sloppy
%\scrollmode
%%%%%%%%%%%%%%%%%%%
\newcommand{\lfrac}[2]{{#1}/{#2}}
\newcommand{\sfrac}[2]{{\small \hbox{${\frac {#1} {#2}}$}}}
\newcommand{\ben}{\begin{eqnarray}}
\newcommand{\een}{\end{eqnarray}}
\newcommand{\la}{\label}
%
%%%%%%%%%%%%%%%%%%
\section{Introduction}
At present general relativity (GR) is the most successful theory
of gravity in describing  gravitational phenomena at
laboratory--, earth-surface--, solar-system-- and star-systems scales
\cite{Will}. Although it gives quite good a description of these
phenomena at galaxies scales and at the scales of the whole
Universe, there exist well known problems such as
the rotation of galaxies, the initial
singularity problem, the early Universe, the recent discovery of the
accelerated expansion of the Universe, etc. 
It is likely, that the indicated problems call for an extension of the
GR framework.

The most promising modern theories which incorporate naturally GR 
like supergravity and (super)string theories, unfortunately introduce,
at least at their present stage of development,
a large number of new fields without any real physical basis
and are far from being  experimentally testable.
Therefore, it seems sensible to explore some other models
which are (i) compatible with the established
gravitational experiments,  
(ii) obtained through a minimal extension of GR,
(iii) promising candidates for
overcoming the above problems, and which 
(iv) may be considered as a part of some more
general modern theories. In the present article we outline the general
properties of one such model which is built upon
an additional scalar field $\Phi$
and that differs from the known inflationary and quintessential models.

We call {\em minimal dilatonic gravity} (MDG) the scalar-tensor
model of gravity \cite{Will}, \cite{BD} with an action
\ben
{\cal A}_{G,\Lambda}=-{\frac c {2\bar\kappa}}\int d^4x
\sqrt{|g|} \Phi \bigl( R + 2 \Lambda \Pi(\Phi) \bigr).
\la{A_Gc}
\een
Here, the well known Brans-Dicke factor has been denoted by
$\omega(\Phi)\!\equiv\!0$, $\Lambda $ stays for the cosmological
constant, while $\Pi(\Phi)$ is the dimensionless cosmological 
factor. The matter action, ${\cal A}_M$, and the matter equations
of motion will have the usual GR form and {\em do not dependend} 
directly on the dilaton field $\Phi$.

The field equations for the metric $g_{\alpha\beta}$ and the
dilaton field $\Phi$ are given by:
\ben
\Phi
\left(G_{\alpha\beta}\!-\!\Lambda \Pi(\Phi) g_{\alpha\beta}\right)
\!-\!(\nabla_\alpha \nabla_\beta\!-\!g_{\alpha\beta}{\Box})\Phi
\!=\! {\sfrac {\bar\kappa} {c^2}} T_{\alpha\beta}, \nonumber \\
{\Box}\Phi\!+\!\Lambda {\sfrac {dV}{d\Phi}}(\Phi)=\! {\sfrac
{\bar\kappa} {3 c^2}}T
\la{FEq}
\een
and yield the  energy-momentum conservation law,
$\nabla_\alpha\,T^\alpha_\beta=0$.

The variation of the action (\ref{A_Gc}) with respect to the metric
$g_{\alpha\beta}$ lieds to the first one of the equations (\ref{FEq}).
The variation of (\ref{A_Gc}) with respect to the dilaton field $\Phi$
produces the following local algebraic relation between this field
and the scalar curvature $R$:
\ben
R+2\Lambda \bigl(\Phi{\sfrac{d\Pi}{d\Phi}}(\Phi)
+\Pi(\Phi)\bigr)=0.
\la{RV}
\een
In combining the latter relation with the trace of the first of 
the equations (\ref{FEq}), the field equation for the
dilaton $\Phi$ is obtained.
Because of the absence of the standard kinetic term
$\sim\omega(\Phi)(\nabla \Phi)^2$ for the dilaton in action
(\ref{A_Gc}), its dependence on nonzero space-time curvature
($R\,\, {/ \hskip -.33cm\equiv}\, 0$) is essential for the
existence of this equation. There, ${\sfrac {dV}{d\Phi}}={\sfrac 2
3} \Phi\left(\Phi{\sfrac{d\Pi}{d\Phi}}-\Pi\right)$ is the derivative
of the dilatonic potential $V(\Phi)$.

The presence of a potential term for the scalar fields 
in the action of the theory, 
was considered many times from different points of view. 
For example, the extended inflation models 
and the quintessential models contain
such a term, see \cite{Steinhard} and the references therein.

In Brans-Dicke's theory, $\Pi(\Phi)\equiv 0$, and one has to make a 
choice for the function $\omega(\Phi)$ only, in order to fix the model. 
However, recent astrophysical data from type Ia supernovae at high red-shifts, 
on the one hand, as well as from cosmic microwave
background, on the other hand, 
lean a strong support  
to  the presence of an additional $\Lambda$-term in the
action of the theory \cite{ICHEP}.

In more general scalar-tensor theories of gravity, two unknown
functions,  $\omega(\Phi)$ and $\Pi(\Phi)$, are to be fixed. The 
old attempts to extract 
from the known gravitational experiments
simultaneously with the dilaton mass  $m_\Phi\neq 0$
also a constant parameter $\omega =const\neq 0$, 
turned out to be less successful, living great uncertainties for the
corresponding values of the above parameters, see \cite{STGExp}. 
A new promising method to
determine $\omega(\Phi)$ and $\Pi(\Phi)$ (or some
equivalent functions) using astrophysical observations,
was developed in the recent work \cite{StarobinskyA}. Unfortunately,
at present, we don't have sufficiently reliable data,
as required for applying this method, so that, at present, 
the form of both functions still remains  an open problem.

As a scalar-tensor theory with only one unknown function
$\Pi(\Phi)$, the MDG is much better defined. Its investigation was
pioneered  by O'Hanlon in connection with Fujii's theory of massive
dilatons as early as in \cite{Fujii}, however, 
without any connection with the cosmological constant problem.

In general, such a minimal model may be related with the so called
nonlinear theories of gravity (see \cite{NLG} and the references
therein). Indeed, the variation of the action (\ref{A_Gc}) with
respect to the dilaton field $\Phi$  gives the local {\em
algebraic} relation (\ref{RV}) between scalar curvature $R$
and dilaton $\Phi$ in place of the {\em differential} equation, 
which is due to the absence of the 
Brans-Dicke kinetic term $\sim\omega(\Phi)(\nabla
\Phi)^2$ in the action for $\omega \equiv 0$. If ${\sfrac
{d^2}{d\phi^2}} \left(\Phi \Pi(\Phi)\right)\,\, {/ \hskip
-.33cm\equiv}\, 0$, one can solve the relation (\ref{RV}) with
respect to the field $\Phi$ and it becomes a {\em local} function
of the scalar curvature: $\Phi= \Phi(R)$. (For $\omega\neq 0$ the
last relation will have non local integral form.) The substitution
of this function back into the action (\ref{A_Gc}) transforms it
to the action of nonlinear theories of gravity ${\cal A}_{NG}=
-{\sfrac c {2\kappa}}\int d^4 x \sqrt{|g|}\,f(R)$ with 
$f(R) = R\Phi(R) + 2\Lambda \Phi(R)\Pi(\Phi(R))$. Hence, the
action (\ref{A_Gc}) is of the Helmholtz type for nonlinear
gravity \cite{NLG}.

MDG may be considered also as a part of more general theories of
gravity like the (4+1)-dimensional Kaluza-Klein model described by
Fujii in \cite{Fujii2}, or, the model of gravity with violated local
conformal symmetry in affine-connected spaces, which probably may
be related to string theory \cite{FF}.

In this special scalar-tensor model of gravity the cosmological
factor $\Pi(\Phi)$ is the only function which has to be chosen to
comply with gravitational experiments and astrophysical
observations. In the present article, we shall demonstrate that 
this results in much better defined
predictions for the dilaton mass, which is now derived from
known gravitational experiments. It turns out that the results of
these experiments on Earth surface and in solar, or star systems
depend mainly on the mass term in the dilaton potential. A numerical
analysis of the boson stars in massive dilatonic gravity based on
MDG was performed very recently \cite{Fiziev2}. It shows that the
boson star structure is sensitive, typically within a few percent
accuracy only, to the dilaton mass rather than to the exact form of 
the potential. Hence, we need some other source of information about
the form of the cosmological factor.

In the present article we shall show how one can solve {\em the
inverse cosmological problem} in MDG. By inverse cosmological
problem we mean the determination of the cosmological factor
$\Pi(\Phi)$ which yields a given time evolution of the scale
parameter $A(t)$ in the Robertson-Walker (RW) model of the
Universe. Such an approach for the determination of the
cosmological factor may be considered as a further development of 
the general idea to find the dilaton potential from astrophysical 
observations as discussed first in the articles \cite{Starobinsky}. 
There, the relation between the dilaton potential and
the luminosity distance was obtained from observations of 
type Ia supernovae explosions at high red-shifts. 
Similar consideration based on real observational data may take place 
in the model of MDG, too. Our approach, in being based upon the evolution 
of the cosmological factor,
gives another possibility to extract the dilaton potential from data
and seems to be more profound.

An essential new element of our MDG is the {\em nonzero} constant $\Lambda$
which we identify with standard cosmological constant.
The new astrophysical data $\Omega_{\Lambda}\!=\!.65\pm .13$,
$H_0\!=\!(65\pm 5)\,km\,s^{-1}Mps^{-1}$ \cite{ICHEP}
give observed value $\Lambda^{obs}=3\Omega_\Lambda H_0^2 c^{-2}=
(.98 \pm .34) \times 10^{-56} cm^{-2}$.
Though the present confidence level of $\Omega_\Lambda$ is not as high,
we accept its value as a basic quantity which defines the natural units
for all cosmological quantities: the cosmological length
$A_c = 1/\sqrt{\Lambda^{obs}} = (1.02\pm .18)\times 10^{28} cm$,
the cosmological time $T_c:=A_c/c= (3.4\pm .6)\times 10^{17} s=
(10.8 \pm 1.9) Gyr$, the cosmological energy density
$\varepsilon_c:={\frac {\Lambda c^2} {\kappa}}=
(1.16\pm .41)\times 10^{-7} g~cm^{-1}~s^{-2}$,
the cosmological energy $E_c:=3 A_c^3\varepsilon_c=
3\Lambda^{-1/2} c^2 \kappa^{-1} =(3.7\pm .7) \times 10^{77} erg$
the cosmological momentum
$P_c:= 3c / (\kappa \sqrt{\Lambda^{obs}}) =
(1.2 \pm .2)\times 10^{67} g~cm~s^{-1}$
and new the cosmological unit for the action
${\cal A}_c:=
3c/(\kappa\Lambda^{obs})=(1.2 \pm .4)\times 10^{122}\,\hbar$,
$\kappa$ being Einstein constant.
Further, we use dimensionless variables like:
$\tau := t/T_c$, $a := A/A_c$, $h:= H\,T_c$
($H:=A^{-1}dA/dt$ being Hubble parameter),
$\epsilon_c = \varepsilon_c / |\varepsilon_c|=\pm 1$,
$\epsilon:=\varepsilon / |\varepsilon_c|$-matter energy density, etc.

\section{Solar System and Earth-Surface Gravitational Experiments}

We use the well studied scalar-tensor theories of gravity
\cite{Will}, \cite{BD}, \cite{STGExp}, \cite{Fujii} to derive the
properties of cosmological factor $\Pi(\Phi)$ and predictions for
gravitational experiments.

\subsection{General Consideration}

1. The MDG with $\Lambda=0$ is a Brans-Dicke theory with $\omega=0$
and contradicts the data from the solar system gravitational experiments,
because the value ${\sfrac 1 2}$ of the post
Newtonian parameter $\gamma = {\frac {1+\omega} {2+\omega}}$ is far from
the experimental constraint $\gamma=1 \pm .001$ \cite{Will}.
We must introduce cosmological term $\Lambda\Pi(\Phi) \neq 0$
in action (\ref{A_Gc}) to overcome this problem.

2. In contrast to O'Hanlon's model \cite{Fujii} we wish MDG to reproduce
GR with $\Lambda\!\neq\!0$ for $\Phi\!=\!\bar\Phi\!=\!const\neq\!0$.
Then from action (\ref{A_Gc}) we obtain normalization condition
$\Pi(\bar\Phi)=1$ and Einstein constant $\kappa=\bar\kappa/\bar\Phi$.

3. In vacuum, far from matter MDG have to allow weak field
approximation $\Phi=\bar\Phi(1+\zeta)$,\,\,$|\zeta|\ll 1$ which we
consider in harmonic gauge. Then the second of equations
(\ref{FEq}), written in linearized form with respect to $\zeta$:
$$\Box \zeta+ \zeta/l_\Phi^2\!=\!{\sfrac {\kappa}{3 c^2}}T,$$
gives ${\sfrac{d\Pi}{d\Phi}}(\bar\Phi)=1/\bar\Phi$. Taylor series
expansion of the function ${\sfrac{dV}{d\Phi}}(\Phi)$ around the
value $\bar\Phi$ starts with term which can be connected with the
dilaton mass $m_{\Phi}={\sfrac {h} {l_\Phi c}}$, $l_\Phi$ being the
dilaton Compton length. Thus we obtain $ {\sfrac
{d^2\Pi}{d\Phi^2}}(\bar\Phi)\!=\!{\sfrac 3 2}p^{-2}\bar\Phi^{-2}$
and
\ben
\Pi\!=\!1\!+ \zeta\!+{\frac 3 {4 p^2 }}\zeta^2\!+O(\zeta^3),
\la{Pi}
\een
$p\!=\!{\sfrac {l_\Phi}{A_c}}$ being a new dimensionless parameter.

4. For few point particles of masses $m_a$
being source of metric and dilaton fields in equations (\ref{FEq})
we obtain Newtonian approximation
($g_{\alpha\beta}\!=\!\eta_{\alpha\beta}\!+\!h_{\alpha\beta}$,
$\!|h_{\alpha\beta}|\!\ll\!1$) for the gravitational potential
$\varphi({\bf r})$ and the dilaton field $\Phi({\bf r})$:
\ben
\varphi({\bf r})/c^2\!=\!
- {\sfrac G {c^2}}\!\sum_a\!{\sfrac {m_a}{|{\bf r - r}_a|}}\!
\left(\!1\!+\!\alpha(p) e^{-|{\bf r - r}_a|/l_\Phi} \right)
\nonumber\\
- {\sfrac 1 6}p^2
\sum_a\!{\sfrac{m_a} M}\left(|{\bf r - r}_a|/l_\Phi\right)^{\!2},
\nonumber \\
\!\Phi({\bf r})/\bar\Phi\!=\!1+\!{\sfrac 2 3}
{\sfrac G{c^2(1-{\sfrac 4 3} p^2)}}
\sum_a\!{\sfrac {m_a}{|{\bf r - r}_a|}}
e^{- |{\bf r - r}_a|/l_\Phi},
\la{SolNewton}
\een
$G\!=\!{\sfrac {\kappa c^2} {8\pi}}(1\!-\!{\frac 4 3}p^2)$ is
Newton's constant, $M\!=\!\sum_a m_a$. The term $$-{\sfrac 1 6}p^2\!
\sum_a{\sfrac{m_a} M}\left(|{\bf r\!-\!r}_a|/l_\Phi\right)^2\!=\!
-{\sfrac 1 6} \Lambda|{\bf r}-\!\sum_a \!{\sfrac {m_a} M}{\bf
r}_a|^2\!+\!const$$ in $\varphi({\bf r})$ is known from GR with
$\Lambda \neq 0$. It represents an universal anty-gravitational
interaction of test mass $m$ with any mass $m_a$ via repellent
elastic force
\ben
{\bf F}_{\!{}_\Lambda\,a}={\sfrac 1 3}\Lambda m c^2 {\sfrac{m_a} M}({\bf r - r}_a).
\la{Fel}
\een

For solar system distances $l\leq 1000 AU$ the whole second term
in $\varphi$ may be neglected \cite{SSL}, being of order $\leq
10^{-24}$. Then we arrive to the known form of gravitational
potential $\varphi$ \cite{STGExp}, but with specific for MDG
coefficient $$\alpha(p)={\sfrac{1+4p^2}{3-4 p^2}}$$ and the
comparison of both possibilities: $\alpha \geq\!{\sfrac 1 3}$, or
$\alpha \leq -1$ with Cavendish type experiments yields the
experimental constraint $l_\Phi \leq 1.6$ [mm] ($p < 2\times
10^{-29}$), see the articles by De~R\'ujula and by Fischbach and
Talmadge in \cite{STGExp} and the references therein. Hence, in
the solar system phenomena the factor $e^{-l/l_\Phi }$ has
fantastically small values ($<exp(-10^{13})$ for the Earth-Sun
distances $l$, or $< exp(-3\times 10^{10})$ for the Earth-Moon
distances $l$) and there is no hope to find some differences
between MDG and GR in this domain.

The constraint $m_\Phi c^2 \geq 10^{-4}[eV]$ does not exclude
a small value (with respect to the elementary particles scales)
for the rest energy of the hypothetical $\Phi$-particle.

5. The parameterized-post-Newtonian (PPN) solution of equation (\ref{FEq})
is complicated, but because of the constraint $p < 10^{-28}$
we can neglect the second term in $\varphi$,
set $\alpha\!=\!{\sfrac 1 3}$ and use with great precision Helbig's
PPN formalism which differs essentially from the standard one \cite{Will}
for zero mass dilaton fields. We obtain much more definite predictions then
general relations between $\alpha$ and the length
$l_\Phi$ given in the articles by De~R\'ujula and by Helbig \cite{STGExp},
because of the condition $\omega\equiv 0$.

\subsection{The Basic Gravitational Effects in MDG}

A brief treatment of basic gravitational effects gives:

\subsubsection{Nordtvedt Effect}

In MDG a body with significant gravitational self-energy
$E_{{}_G}=\sum_{b\neq c} G{\sfrac {m_b m_c}{|{\bf r}_b - {\bf r}_c|} }$
will not move along
geodesics due to additional universal {\em anty-gravitational} force:
\ben
{\bf F}_{\!{}_N} = -{\sfrac 2 3} E_{{}_G}\nabla \Phi.
\la{NordtF}
\een
For usual bodies it is too small even at distances
$|{\bf r-r}_a|\!\leq\!l_\Phi$, because of the small factor $E_G$.
Hence, in MDG we have no strict strong equivalence principle. Nevertheless,
the week equivalence principle is not violated.

The experimental data for Nordtvedt effect caused by the Sun are
formulated as a constraint $\eta=0 \pm .0015$ \cite{Will} on the
parameter $\eta$ which in MDG becomes a function of the distance
$l$ to the source:
$$\eta(l)=-{\sfrac 1 2}\left(1+l/l_\Phi\right)
e^{-l/l_\Phi}.$$
Taking into account the value of the Astronomical
Unit (AU) $l_{{}_{AU}} \approx 1.5\times 10^{11}[m]$ we obtain
from the experimental value of $\eta$ the constraint $l_\Phi \leq
2\times 10^{10 }[m]$.

\subsubsection{Time Delay of Electromagnetic Waves}

The standard action for electromagnetic field and the Maxwell
equations do not depend directly on the field $\Phi$. Therefore
the influence of this field on the electromagnetic phenomena like
motion of electromagnetic waves in vacuum is possible only
indirectly -- via its influence on the space-time metric. The
solar system measurements of the time delay of the electromagnetic
pulses give the above used value  $\gamma=1 \pm .001$ of this post
Newtonian parameter \cite{Will}. In MDG this yields the relation
$(1 \pm .001) g( l_{{}_{AU}})= 1$ and gives once more the
constraint $l_\Phi \leq 2\times 10^{10 }[m]$. Here
$$g(l):=1+{\sfrac 1 3 }(1+l/l_\Phi) e^{-l/l_\Phi}.$$

\subsubsection{Perihelion Shift}

Helbig's results \cite {STGExp} applied in MDG give the following
formula for the perihelion shift of a planet orbiting around the
Sun (with mass $M_\odot$): $$\delta \varphi = {\frac
{k(l_p)}{g(l_p)}}\delta\varphi_{{}_{GR}}.$$ Here $l_p$ is the
semi-major axis of the orbit of planet and $$k(l_p) \approx 1 +
{\sfrac 1 {18}}\left( 4 + {\sfrac{l_p^2}{l_\Phi^2}}{\sfrac {l_p
c^2}{GM_\odot}}\right) e^{-l_p/l_\Phi} -{\sfrac 1 {27}}
e^{-2l_p/l_\Phi}$$ is obtained neglecting its eccentricity. The
observed value of perihelion shift of Mercury gives the constraint
$l_\Phi \leq  10^{9 }[m]$ (see the article by De~R\'ujula \cite
{STGExp}).

The above week restrictions on $l_\Phi$ derived in MDG from gravitational
experiments show that in presence of dilaton field $\Phi$
are impossible essential deviations from GR in solar system.
Formulae (\ref{SolNewton}) show that observable deviations from Newton law
of gravity can not be expected at distances greater then few cm, too.

\section{Application of MDG  in Cosmology}

\subsection{General Equations of MDG for RW Universe}

The design of {\em realistic} model of the Universe lies beyond
the scope of the present article. Here we wish only to outline
some general features of the MDG, applied to cosmological problems
and to show that in general it is able to correct the well known
shortcomings of GR in this domain. We derive the equations of the
inverse cosmological problem in MDG and demonstrate some simple
solutions, which give indications for unexpected new physics.

Consider RW adiabatic homogeneous isotropic Universe with
$ds^2_{RW}= c^2 dt^2 - A^2 dl^2_k$, $t=T_c\tau$,
$A(t)=A_c a(\tau)$ and dimensionless
$dl^2_k={\frac {dl^2}{1-kl^2}+l^2(d\theta^2+sin^2\theta)d\varphi^2}$
($k=-1, 0, 1$) in presence of matter with energy-density
$\varepsilon=\varepsilon_c\epsilon(a)/\bar\Phi$ and
pressure $p=\varepsilon_c p_\epsilon(a)/\bar\Phi$.
%allowing only adiabatic processes.

From equations (\ref{FEq}) and (\ref{RV}) we obtain the following
basic dynamical equations of MDG for RW Universe (see for details
the Appendix):
\ben
{\sfrac 1 a}{\sfrac {d^2\!a}{d\tau^2}}+
{\sfrac 1 {a^2}}({\sfrac {da}{d\tau}})^{{}_2}+{\sfrac k {a^2}}=
{\sfrac1 3} \left(\Phi{\sfrac{d\Pi}{d\Phi}}(\Phi)+\Pi(\Phi)\right),\nonumber\\
{\sfrac 1 a}{\sfrac {da}{d\tau}}{\sfrac {d\Phi}{d\tau}}+
\Phi \left({\sfrac 1 {a^2}}({\sfrac {da}{d\tau}})^{{}_2}
+{\sfrac k {a^2}}\right)={\sfrac1 3}\left(\Phi\Pi(\Phi)+ \epsilon(a)\right).
\la{DERWU}
\een

Using Hubble parameter
$h(a)=a^{-1}{\sfrac {da}{d\tau}}(\tau(a))$ ($\tau(a)$
is the inverse function of $a(\tau)$), new variable $\lambda=\ln a$ and
denoting by prime the differentiation with respect to $\lambda$
we write down the equations (\ref{DERWU}) as a second order
non-autonomous system for the functions $\Phi(\lambda)$ and $h^2(\lambda)$:
\ben
{\sfrac 1 2}(h^2)^\prime +2 h^2 + k e^{-2\lambda}=
{\sfrac 1 3}\left(\Phi{\sfrac{d\Pi}{d\Phi}}(\Phi)+\Pi(\Phi)\right),\nonumber \\
h^2 \Phi^\prime +\left(h^2+k e^{-2\lambda}\right)\Phi=
{\sfrac 1 3}(\Phi\Pi(\Phi)+\epsilon(e^\lambda)).
\la{NDE}
\een
For given function $\Pi(\Phi)$ the solutions of system (\ref{NDE})
determine the functions $\Phi(a)$ and $h(a)$ and the
time dependence of the scale parameter $a(\tau)$ 
is given implicitly by the relation
\ben
\tau(a)=\int^a_{\!a_{in}}\!da /(a\,h(a))+\tau_{in}.
\la{tau}
\een
Instead, we can exclude the cosmological factor $\Pi(\Phi)$
arriving to {\em linear} differential equation of second order for
the function $\Phi(\lambda)$:
\ben
\Phi^{\prime\prime}\!+\!
\left({\sfrac {h^{\prime}} h}\!-\!1\right)\Phi^{\prime}
\!+\!2\left({\sfrac {h^{\prime}} h}\!-
k h^{-2}e^{-2\lambda}\right)\Phi
\!=\!{\sfrac 1 {3h^2}}\epsilon^{\prime}.
\la{ODEPhi}
\een
In terms of the function $\psi(a) = \sqrt{|h(a)|/a}\,\Phi(a)$ it reads:
\ben
\psi^{\prime\prime} + n^2 \psi = \delta,
\la{DEPsi}
\een
where we have introduced the new functions
\ben
-n^2 = {\sfrac 1 2}{\sfrac {h^{\prime\prime}} h}-
{\sfrac 1 4}({\sfrac {h^{\prime}} h})^2\!-
{\sfrac 5 2}{\sfrac {h^{\prime}} h}+{\sfrac 1 4}
+{\sfrac {2 k} {h^2}} e^{-2\lambda},\nonumber \\
\delta = {\sfrac 1 3}\sqrt{{ a/{|h|^3}}}
{\sfrac {d\epsilon}{da}} .
\la{n,delta}
\een

For the important coefficient $n$  one can obtain the
representation $$-n^2 = 3+{\sfrac{2 k}{a^2 h^2}}+{\sfrac S {2
h^2}}+ {\sfrac 5 2}q $$ where $-q={\sfrac 1
{ah^{2}}}{\sfrac{d^2a}{d\tau^2}}= 1+{\sfrac {h^{\prime}} h}$ is
decelerating parameter  and
$S[a(\tau)]\!=\!{\sfrac{d^3a}{d\tau^3}}/{\sfrac{da}{d\tau}}-
{\sfrac 3
2}\bigl({\sfrac{d^2a}{d\tau^2}}/{\sfrac{da}{d\tau}}\bigr)\!{}^2$
is Schwartz derivative of the scale factor $a(\tau)$.

\subsection{The Inverse Cosmological Problem}

Now we are ready to consider the inverse cosmological problem: to
find the corresponding cosmological factor $\Pi(\Phi)$ and
dilatonic potential $V(\Phi)$ which leads to a  given evolution of the
Universe, determined by function $a(\tau)$. It is remarkable that
in MDG an unique solution of this problem exist for almost any
three times differentiable function $a(\tau)$. Indeed, for given
$a(\tau)$ we may construct a function $h(\lambda)$ and find the
general solution $\Phi(\lambda,C_1,C_2)$ of the linear second
order differential equation (\ref{ODEPhi}). It depends on two
arbitrary constants $C_{1,2}$ which have to be determined from the
additional conditions\, $\Pi(\bar\Phi)=1,\,\, {\sfrac
{d\Pi}{d\Phi}}(\bar\Phi)=\bar\Phi^{-1},\,\, {\sfrac
{d^2\Pi}{d\Phi^2}}(\bar\Phi)={\sfrac 3 2}p^{-2}\bar\Phi^{-2}$.
They ensure the self-consistence of the MDG model yielding initial
conditions for equation (\ref{ODEPhi}) at point $\bar\lambda$
which is real solution of the algebraic equation
\ben
r(\bar\lambda)=-4.
\la{r=-4}
\een
This is a new form of relation (\ref{RV}) and
$r(\lambda)=-6\left({\sfrac 1 2}(h^2)^\prime +2 h^2 +k
e^{-2\lambda}\right)$ is the dimensionless scalar curvature $r=
R/\Lambda$. Then:
\ben
\bar\Phi=-4\bar\epsilon \left(1\!+\!{\sfrac 4 3}p^2\right)/
\!\left(\bar j_{00}^\prime(1\!+\!{\sfrac 4 3}p^2) +
4p^2\bar h^2\bar r^\prime \right), \nonumber \\
\bar \Phi^\prime/\bar \Phi =
-{\sfrac 1 3}p^2 \bar r^\prime/\left(1\!+\!{\sfrac 4 3}p^2\right).
\la{barPhiPhi'}
\een
Here and later on, the bar sign means that the 
corresponding quantities are
calculated at the point $\bar\lambda$, and
$j_{00}=G_{00}/\Lambda=3\left(h^2+ke^{-2\lambda}\right)$
is the dimensionless $00$-component of Einstein tensor.
Hence, the values of all "bar" quantities
(including $\bar \kappa$ in action (\ref{A_Gc}))
may be determined from the time evolution $a(\tau)$ of the Universe via
the solution $\bar \lambda = \ln \bar a$ of the equation (\ref{RV}).

In their turn the quantities $\bar\Phi$ and $\bar\Phi^\prime$
determine the values of the constants $C_{1,2}$ and an unique solution
$\Phi(\lambda)$ of the equation (\ref{ODEPhi}). Indeed, let
$\Phi_1(\lambda)$ and $\Phi_2(\lambda)$ are fundamental system of
solutions of the homogeneous equation associated with
non-homogeneous one (\ref{ODEPhi}). Then $\Delta(\lambda):=
\Phi_1\Phi_2^\prime - \Phi_2\Phi_1^\prime= (\bar\Delta \bar h/\bar
a)e^{\lambda}\,h(\lambda)\neq 0$ and
\ben
\Phi(\lambda)= C_1\Phi_1(\lambda) + C_2\Phi_2(\lambda)+
\Phi_\epsilon(\lambda)
\la{sol}
\een
is the general solution of equation (\ref{ODEPhi}) which depends
on two constants
$$C_1=(\bar\Phi_2^\prime\bar\Phi-\bar\Phi_2\bar\Phi^\prime)/\bar\Delta,\,\,\,\,
C_2=(\bar\Phi_1\bar\Phi^\prime-\bar\Phi_1^\prime\bar\Phi)/\bar\Delta$$
and on contribution of matter described by the term:
$$\Phi_\epsilon\!=\!\bar a/(3\bar h\bar\Delta)\!\left(
\Phi_2\!\int_{\bar\lambda}^\lambda\!d\epsilon\,\Phi_1/(ah)-
\Phi_1\!\int_{\bar\lambda}^\lambda\!d\epsilon\,\Phi_2/(ah)\right).$$

The cosmological factor $\Pi$ and the potential $V$
as functions of the variable $\lambda$ are determined by the equations
\ben
\Pi(\lambda) = j_{00} + 3 h^2 \Phi^\prime/\Phi -
\epsilon/\Phi,\,\,\,\,\, V(\lambda)={\sfrac 2 3}
\int\Phi\left(\Phi\Pi^\prime-\Phi^\prime\Pi \right)d\lambda
\la{PiV}
\een
which define the functions $\Pi(\Phi)$ and $V(\Phi)$ implicitly, as well.

\subsection{Simple Examples}

Let us demonstrate the above procedure for solution of the inverse
cosmological problem on several simple examples:

1. Evolution law $a(\tau)\!=\!(\omega\tau)^{1/\nu}$, (where
$\omega$ is a free parameter) gives $$h(\lambda)\!=\!{\sfrac \omega
\nu}e^{-\nu\lambda},\,\,\,\,\,\,q=\nu-1,\,\,\,\,\,\,
-n^2(\lambda)=
{\sfrac{1+10\nu+\nu^2}4}+2k{\sfrac{\nu^2}{\omega^2}}e^{2(\nu-1)\lambda}$$
 and the equation
$$\bar a^2\!+{\sfrac {3(\nu-2)\omega^2}{2\nu^2}}\bar
a^{2(1-\nu)}\!=\!k$$ for $\bar a$. Then:

i) For $\nu\geq 2$ we have real solution $\bar a$ only if $k=+1$:
$$ \Phi_1(a)=a^{{\frac {\nu+1}2}}I_\mu\!\left(b
a^{\nu-1}\right),\,\,\,\,\, \Phi_2(a)=a^{{\frac
{\nu+1}2}}K_\mu\!\left(b a^{\nu-1}\right), $$ $\mu:={ \sfrac
{\sqrt{1+10\nu+\nu^2}} {2(\nu-1)} } $ being the order of Bessel
functions $I_\mu,J_\mu,K_\mu,Y_\mu$ and $b:=\sqrt{{\sfrac
{2\nu^2}{(\nu-1)^2\omega^2}}}$. For GR-like law
$a\!\sim\!\sqrt{\tau}$ ($\nu\!=\!2$) in MDG we obtain positive
value ($k=+1$) for the three-space curvature, $\bar\lambda\!=\!0$,
$\mu={\sfrac 5 2}$ and Bessel functions are reduced to elementary
functions.

ii) When $\nu<2$ all values $k=-1,0,+1$ are admissible:

- for $k\!=\!+1$ the solutions $\Phi_{1,2}$ are the same as above;

- for $k\!=\!0$ we have $$\Phi_{1,2}=a^{{\frac
{\nu+1}2}\pm\mu(\nu-1)};$$

- for $k\!=\!-\!1$ the solutions are: $$ \Phi_1\!=\!a^{{\frac
{\nu+1}2}}J_\mu\!\left(ba^{\nu-1}\right),\,\,\,\,\,\,\,
\Phi_2\!=\!a^{{\frac {\nu+1}2}}Y_\mu\!\left(ba^{\nu-1}\right). $$
In the special case of linear evolution $a\!\sim\!\tau$ ($\nu=1$)
$-n^2(\lambda)=3+{\sfrac {2k}{\omega^2}}$,\,
$$\Phi_{1,2}(a)=a^{1\pm \sqrt{-n^2}}$$ and the root
$\bar\lambda\!=\!{\sfrac 1 2}\ln{\sfrac 3 2}(k+\omega^2)$ is real
for all values of $\omega^2\!>\!0$, if $k\!=\!0,+1$. For
$k\!=\!-1$ the root $\bar\lambda$ will be real if
$|\omega|\!>\!1$.

2. Evolution law $a(\tau)\!=\!e^{\omega\tau}$ gives
$$h(\lambda)\!=\!\omega,\,\,\,\,\,\,\,q=-1,\,\,\,\,\,\,\,-n^2(\lambda)\!=\!{\sfrac
1 4}+{\sfrac {2k}{\omega^2}}e^{-2\lambda}$$ and the equation
${\sfrac 2 3}\bar a^2 (1-3\omega^2)\!=\!k$ whith root $\bar
a\!=\!\sqrt{{\sfrac 3 {2|1-3\omega^2|}}}$. Now we have the
following solutions:

i) For  $|\omega|\!<\!\!{\sfrac{\sqrt{3}}3}$ and $k\!=\!+1$:
$$\Phi_1\!=\!a\cosh\!\left(\!{\sfrac{\sqrt{2}}{|\omega|a}}\!\right),\,\,\,\,\,\,\,
\Phi_2\!=\!a\sinh\!\left(\!{\sfrac{\sqrt{2}}{|\omega|a}}\!\right);$$

ii) For $|\omega|\!>\!\!{\sfrac{\sqrt{3}}3}$ and $k\!=\!-1$:
$$\Phi_1\!=\!a\,\cos\!\left(\!{\sfrac{\sqrt{2}}{|\omega|a}}\!\right),\,\,\,\,\,\,\,
\Phi_2\!=\!a\,\sin\!\left(\!{\sfrac{\sqrt{2}}{|\omega|a}}\!\right).$$

Conditions (\ref{r=-4}) and (\ref{barPhiPhi'}) exclude {\em exact}
exponential expansion of spatially flat Universe ($k\!=\!0$ ). For
inflationary scenario in this case one may use a scale factor
$a(\tau)\!=\!e^{\omega\tau}\!+\!const$ which turns out to be possible
if $const\!\neq\!0$.

\section{Some Specific Properties and possible further developments of the MDG
model}

In the following concluding remarks we stress some basic
properties of MDG models and outline some possible important
further developments to be considered elsewhere. First of all
we have to stress that actually here we consider the general
properties of certain class of such models. Depending on the form
of cosmological factor $\Pi(\Phi)$ the concrete representatives of
this class of models of extended gravity will have different
physical properties and consequences, having in the same time some
general features.

As we see, the MDG models of the RW Universe have the following specific
properties:

1) In domains where $n\!>\!0$ the dilatonic field $\Phi(a)$ oscillates;
if $n\!<\!0$ such oscillations do not exist (see equation
(\ref{DEPsi})). To judge whether the oscillating solutions of MDG are
physically acceptable, one has to find their physical consequences
and possible manifestations in the real world.

2) The dilatonic field $\Phi$ can change its sign, i.e. in general
phase transitions of the Universe from gravity ($\Phi\!>\!0$) to
anty-gravity ($\Phi\!<\!0$) and vice-versa are possible in MDG. In
general, the hyper-surfaces $\Phi(x)\!=\!0$ in space-time are
dangerous singular Cauchy surfaces: as seen from first of
equations (\ref{FEq}), initial conditions on them do not determine
second time-derivatives of metrics and evolution. This problem
needs careful investigation, but in the case of the RW Universe
considered here it has a simple positive solution: the second of
equations (\ref{FEq}) and equation (\ref{RV}) instead of the first
of equations (\ref{FEq}) determine the evolution even for initial
condition $\Phi(\tau_0)=0$.

Depending on the form of cosmological factor $\Pi(\Phi)$ in some
of MDG models the above transition from gravity to anty-gravity
are possible, but such new phenomena do not exist in models with
proper cosmological factor. To decide which model describes better
the real Universe we first have to study theoretically the
properties of this rich class of models and then to look for
observational evidences in favor of some of them.

3) In the spirit of Max principle Newton's constant 
depends on the presence of matter:
$G\!\sim\!1/\bar\Phi\!\sim\!1/\bar\epsilon$, 
see equation (\ref{barPhiPhi'}).

4) For simple functions $a(\tau)$ the cosmological factor
$\Pi(\Phi)$ and the dilatonic potential $V(\Phi)$ may show unexpected
catastrophic behavior including terms $\sim\!(\Delta\Phi)^{3/2}$
($\Delta\Phi\!=\!\Phi\!-\!\Phi(\lambda^\star)$) in vicinity of the
critical points $\lambda^\star$:
$\Phi^\prime(\lambda^\star)\!=\!0$ of the projection of analytical
curve $\{\Pi(\lambda),\Phi(\lambda),\lambda\}$ on the plain
$\{\Pi,\Phi\}$. The conditions $\Pi^\prime(\lambda^\star)\!=\!0,
\Pi^{\prime\prime}(\lambda^\star)\!=\!0,
\Pi^{\prime\prime\prime}(\lambda^\star)\!\neq\!0$;
$\Phi^\prime(\lambda^\star)\!=\!0,
\Phi^{\prime\prime}(\lambda^\star)\!\neq\!0$ ensure a behavior
$$\Pi(\Phi)\!=\!e^{\sigma{\star}\,
\Delta\Phi}\left(\Pi^\star\!+\!O_{3/2}(\Delta\Phi)
\!+\!O_2(\Delta\Phi)\right),$$ where
$\sigma(\lambda):=\left(3(1+{\sfrac {h^{\prime}}
h})h^2-\epsilon/\Phi\right)/ \left(j_{00}-\epsilon/\Phi\right)$.

The potential $V(\Phi)$ has the same critical points $\Phi^\star$
as the factor $\Pi(\Phi)$ and in numerical study of above examples
shows the following behavior: a local minima around the point
$\bar\Phi$ and critical points at one, or at both ends of some
interval $[\Phi_{left}^\star, \Phi_{right}^\star]\ni \bar\Phi$. An
interesting open problem is to find the general conditions on the
scale factor $a(\tau)$ which make impossible these catastrophes,
yielding an everywhere analytical cosmological factor $\Pi(\Phi)$
and potential $V(\Phi)$ with desired properties. The existence of
such class of scale factors  $a(\tau)$ is clear: in the direct
cosmological problem we can choose a desired functions $\Pi(\Phi)$ and
$V(\Phi)$ and then we will obtain the corresponding function
$a(\tau)$.

5) Clearly one can construct MDG model of Universe without the typical
for GR initial singularities: $a(\tau_0)=0$ and with any desired
kind of inflation.

6) Because dilaton field $\Phi$ is quite massive, in it will be
stored significant amount of energy. In RW model this energy has a
homogeneous and isotropic distribution in the Universe. An
interesting open question is: may the field $\Phi$ play the role
of a dark matter, or of a dark energy in the Universe?

7) A very important problem is to reconstruct the cosmological
factor $\Pi(\Phi)$ of {\em the real Universe} using a proper
experimental data and the astrophysical observations (see
\cite{Starobinsky} where such problem was considered first in the 
framework of more complicated models with two unknown functions).

8) As a further steeps for a justification of the parameters of the
MDG model and its relevance to the physical reality one should
study the possible consequences of the existence of the dilaton $\Phi$
for binary pulsars (which give precise test of the extended theories
of gravity) and for the curves of the rotation of galaxies which are
not explained by Newtonian theory of gravity and GR.

It's clear that MDG offers new curious possibilities which deserve
further careful investigation.

\vskip .5truecm

{\em Acknowledgments:}

The author is deeply indebted to Professor Y. Fujii for the receiving
a Latex source of Appendix 9 of his book \cite{Fujii2} and the help in
the references, 
to S. Yazadjiev for the fruitful discussions and the help in the references, 
to Professor M. Kirchbach and to Professor S. Petcov for the reading of 
the manuscript and for the useful suggestions
and to the unknown referee for the useful suggestions.

This work was supported by the Bulgarian National Fund for Scientific
Researches, Contr. No. F610/1999 and by the Sofia University Research
Fund, Contr. No. 245/1999 and No. 303/2000.

\section{Appendix}

Here we give a more extended explanation of the derivation of the
basic dynamical equations (\ref{DERWU}) of the MDG for RW Universe.
The first of these equations is a direct representation of the
equation (\ref{RV}) written for the case of RW metric. The second
one represents the (0,0) component of the first equation of the system
(\ref{FEq}) for the same case. The only nontrivial additional
equation which one can derive from these generalized Einstein
equations, considering their (i,i) components, is equivalent to the
second equation of the system (\ref{FEq}). (Note that in the case
of RW metric one obtains the same equations for $i=1,2,3$.) For the RW
Universe this equation for the field $\Phi=\Phi(\tau)$ acquires
the form: \ben \ddot\Phi +3{\sfrac {\dot a} a}\dot\phi +{\sfrac
{dV}{d\Phi}}={\sfrac 1 3}\left(\epsilon - 3p_\varepsilon\right).
\la{A1} \een From the conservation law
$\nabla_\alpha\,T^\alpha_\beta=0$, applied for the RW metric, one
obtains the same relation as in GR: ${d\over {da}}\left(a^3
\epsilon\right)= - 3 p_\varepsilon a^3$. It makes possible to
exclude the pressure $p_\varepsilon$ from equation (\ref{A1}).
Now, just as in GR, it becomes possible to prove that the equation
(\ref{A1}) follows from the basic system of dynamical equations
(\ref{DERWU}): differentiating the second one with respect to the
time $\tau$, using the first one and once more the second one,
after some algebra we derive the equation (\ref{A1}). This means
that any solution of the system (\ref{DERWU}) satisfies the
equation (\ref{A1}) and there is no need to take it into account,
solving dynamical problems for the RW Universe.

One has to stress that the second order dynamical equations for the RW
Universe in MDG: the first of the equations (\ref{DERWU}) and the
equation (\ref{A1}), are equivalent to the Euler-Lagrange
equations for the (dimensionless) action: 
\ben
{\alpha}_{{}_{RW}}=\int d\tau \left(-\Phi a \dot a^2 - a^2 \dot
\Phi \dot a + k a \Phi -{\sfrac 1 3
}a^3\bigl(\Phi\Pi(\Phi)+\epsilon(a)\bigr) \right). \la{RWAction}
\een 
It is obtained from the original action (\ref{A_Gc}) (per
unit volume) substituting in it the RW metric and dropping out the
corresponding common factor with dimension of action (i.e. the
cosmological unit for action ${\cal A}_c$, introduced in Sec.1).
The second of the equations (\ref{DERWU}), being a first order
differential equation, represents the corresponding energy
integral of this Euler-Lagrange system:

\ben \epsilon_{total} =\dot a p_a+\dot\Phi p_\Phi -\left(-\Phi a
\dot a^2 - a^2 \dot \Phi \dot a + k a \Phi -{\sfrac 1 3
}a^3\bigl(\Phi\Pi(\Phi)+\epsilon(a)\bigr) \right) = \nonumber \\ -
a^3\left({\sfrac 1 a}{\sfrac {da}{d\tau}}{\sfrac {d\Phi}{d\tau}}+
\Phi \left({\sfrac 1 {a^2}}({\sfrac {da}{d\tau}})^{{}_2} +{\sfrac
k {a^2}}\right)-{\sfrac 1 3}\left(\Phi\Pi(\Phi)+
\epsilon(a)\right) \right) \equiv 0. 
\la{energy} 
\een 
Just as in GR, it must be zero, 
because of the invariance of MDG under general coordinate
transformations. In the equation (\ref{energy})
$p_a = -2\Phi a \dot a-a^2 \dot \Phi$ and $p_\Phi = -a^2\dot a$
are the corresponding (dimensionless) canonical momenta for the
action (\ref{RWAction}).

\end{document}